\documentclass[12pt]{article}
\usepackage{times}   
\usepackage[singlelinecheck=false]{caption} 
\usepackage{graphicx}

\usepackage{xfrac}
\usepackage{relsize}
\usepackage{amsmath}    
\usepackage{cite}   
\begin{document}

%
%
%
\bigskip \textbf{ \LARGE \\  \mbox{Towards a theory of wavefunction collapse Part 1:}  \\
~
\newline
\noindent
\Large How the Di\'{o}si-Penrose criterion and Born's rule can be derived from semiclassical gravity \\
~
\newline
\noindent
How the Di\'{o}si-Penrose criterion can be relativistically generalised with the help of the Einstein-Hilbert action
\\
\\
}

%
%
%

\noindent \textit{Garrelt Quandt-Wiese}
\footnote{
\scriptsize My official last name is Wiese. For non-official concerns, my wife and I use our common family name: Quandt-Wiese.}

\noindent \textit{\small{Schlesierstr.\,16, 64297 Darmstadt, Germany}}\\ 
\noindent  {\it {\small garrelt@quandt-wiese.de}}\\
\noindent  {\it ~ {\small http://www.quandt-wiese.de}}

%
%
%
\bigskip
\begin{quote}
{\small
A new approach to wavefunction collapse is prepared by an analysis of semiclassical gravity. The fact that, in semiclassical gravity, superposed states must share a common classical spacetime geometry, even if they prefer (according to general relativity) differently curved spacetimes, leads to energy increases of the states, when their mass distributions are different. If one interprets these energy increases divided by Planck's constant as decay rates of the states, one obtains the lifetimes of superpositions according to the Di\'{o}si-Penrose criterion and reduction probabilities according to Born's rule. The derivation of Born's rule for two-state superpositions can be adapted to the typical quantum mechanical experiments with the help of a common property of these experiments. It is that they lead to never more than two different mass distributions at one location referring e.g. to the cases that a particle "is", or "is not", detected at the location. From the characteristic energy of the Di\'{o}si-Penrose criterion, an action is constructed whose relativistic generalisation becomes obvious by a decomposition of the Einstein-Hilbert action to the superposed states. In Part 2, semiclassical gravity is enhanced to the so-called Dynamical Spacetime approach to wavefunction collapse, which leads to a physical mechanism for collapse. 
}
\end{quote}

\bigskip
%
%
%
\section{Introduction}                 
%
%
\label{sec:1}
Gravity is the most often discussed candidate for a physical explanation of wavefunction collapse. The Dynamical Spacetime approach to wavefunction collapse, which is prepared in this publication and developed in Part 2 \cite{P2}, also assumes gravity as the driver of collapse. That gravity could be responsible for collapse was first mentioned by Feynman \cite{Fey-1} in the 1960s, and led to a first vague model formulated by K\'{a}rolyh\'{a}zi \cite{Kar-1,Dio-5}. Concrete gravity-based models were developed by Di\'{o}si \cite{Dio-1} and Penrose \cite{Pen-1} in the 1980s and 1990s. In Di\'{o}si's approach, fluctuations of the gravitational field are the driver of collapse. In Penrose's approach, the uncertainty of location in spacetime, which occurs when superposed states prefer differently curved spacetimes due to different mass distributions, plays the central role, which leads to a fuzziness of energy being responsible for the superposition's decay. Interestingly, the approaches of Di\'{o}si and Penrose predict the same lifetimes of superpositions, which can be determined with a characteristic gravitational energy depending on the mass distributions of the superposed states. This rule of thumb, sometimes referred to as the Di\'{o}si-Penrose criterion, is often used for quantitative assessments of experimental proposals investigating certain properties of wavefunction collapse \cite{GExp-8,GExp-12,GExp-4,GExp-13,GExp-14,GExp-11,Exp-EPR_Grav-1,Dio-5}.  

\bigskip
\noindent
Another starting point for a gravity-based collapse model is semiclassical gravity, in which the gravitational field is not quantised and spacetime geometry is treated classically \cite{NS-3,NS-4}. As a consequence, superposed states must share the same classical spacetime geometry, even if they prefer (according to general relativity) differently curved spacetimes, which is the case when their mass distributions are different. This provokes a competition between the states for the curvature of spacetime. However, this mechanism alone cannot explain collapse, which is known from studies of the Schr\"odinger-Newton equation displaying semiclassical gravity in the Newtonian limit \cite{NS-12,NS-2}.

\bigskip
\bigskip
\bigskip
\bigskip
\noindent
The purpose of this paper is to prepare the derivation of the Dynamical Spacetime approach to wavefunction collapse in Part 2 \cite{P2}. This is carried out by an analysis of semiclassical gravity. The Dynamical Spacetime approach enhances semiclassical gravity by the so-called Dynamical Spacetime postulate, which enables it for an explanation of wavefunction collapse. In \cite{NS}, an overview on the derivation and proposed experimental verification of the Dynamical Spacetime approach is given. 

\bigskip
\noindent
The question of whether the gravitational field must not to be quantised and spacetime geometry can be treated classically, as assumed by semiclassical gravity and the Dynamical Spacetime approach, is still the subject of scientific debate \cite{Pen-4,NS-6} and has not been decided by experiments so far \cite{NS-15}. 

\bigskip
\noindent
The analysis of semiclassical gravity in this paper will show that there is a relation between semiclassical gravity and the Di\'{o}si-Penrose criterion. This relation will give us an idea of how the Di\'{o}si-Penrose criterion can be relativistically generalised with help of the Einstein-Hilbert action. The analysis of semiclassical gravity will also show that there is possibly a relationship between semiclassical gravity and Born's rule, and that the fact that all the experiments performed so far behave in accordance with Born's rule is related to a property that these experiments have in common. 

\bigskip
\begin{center}
---
\end{center}

\newpage
\noindent
The remainder of this paper is structured as follows. In Section \ref{sec:2}, we recapitulate the Di\'{o}si-Penrose criterion and the approaches of Penrose and Di\'{o}si. In Section \ref{sec:3}, we start with the analysis of semiclassical gravity and show how the Di\'{o}si-Penrose criterion and Born's rule can be derived for two-state superpositions. In Section \ref{sec:4}, we show how the Di\'{o}si-Penrose criterion can be relativistically generalised with the help of the Einstein-Hilbert action. In Section \ref{sec:5}, we show how our derivation of Born's rule for two-state superpositions can be generalised for the typical quantum mechanical experiments, and give a preliminary explanation as to why all experiments so far behave in accordance with Born's rule.

\newpage
%
%
%
\section{Approaches of Penrose and Di\'{o}si}                 
%
%
\label{sec:2}
In this section, we recapitulate the approaches of Penrose and Di\'{o}si. We begin with the Di\'{o}si-Penrose criterion for estimating the lifetimes of superpositions.

%
\begin{figure}[b]
\centering
\includegraphics[width=16cm]{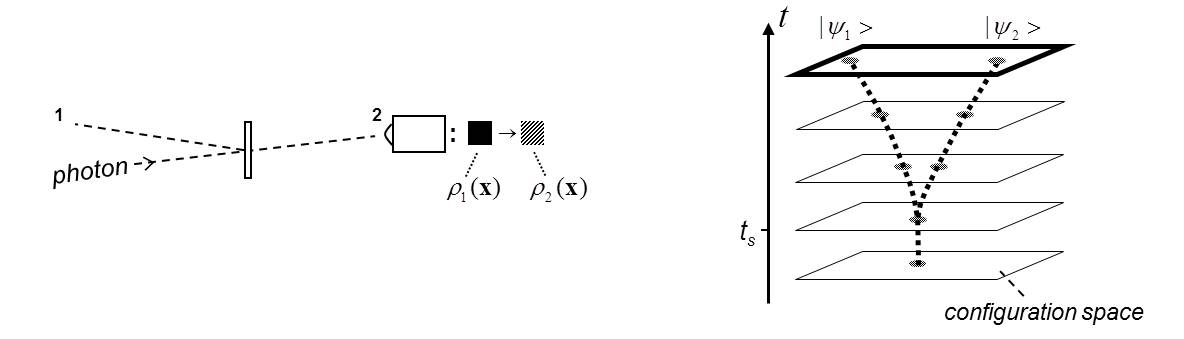}\vspace{0cm}
\caption{\footnotesize
\textit{Left:} \,Experiment to generate a superposition of states with mass distributions $\rho_{_{1}}(\mathbf{x})$ and $\rho_{_{2}}(\mathbf{x})$.\\
\hspace*{19mm} The detector displaces the rigid body for photon detection.\\ 
\hspace*{10mm} \textit{Right:} \,Illustration of the experiment's state vector's evolution in configuration space, which splits\\
\hspace*{21mm} into two wavepackets at {\small$t_{_{s}}$} when the photon enters the beam splitter.
}
\label{fig1}
\end{figure}

\bigskip
%
%
%
\subsection{Di\'{o}si-Penrose criterion}                 
%
%
\label{sec:2.1}
The Di\'{o}si-Penrose criterion is an easy-to-use rule of thumb for estimating the lifetimes of quantum superpositions. The lifetime of a superposition depends on how much the mass distributions of its states differ from each other. A superposition of two states with mass distributions $\rho_{_{1}}(\mathbf{x})$ and $\rho_{_{2}}(\mathbf{x})$ can be generated with the single-photon experiment in the left-hand side of Figure \ref{fig1}, in which the detector displaces a rigid body for photon detection. The mean lifetime of the superposition depends on a characteristic gravitational energy $E_{_{G12}}$, which we call the \textit{Di\'{o}si-Penrose energy}. This energy divided by Planck's constant can be thought of as a decay rate leading to the following lifetime $T_{_{G}}$ of the superposition \cite{Pen-1,Dio-3}:

\begin{equation}
\label{eq:1}
T_{_{G}}\approx\frac{\hbar}{E_{_{G12}}}
 \textrm{\textsf{~~.~~~~~~~~~~~~~~\footnotesize \textit{Di\'{o}si-Penrose criterion}}}
\end{equation}

~
\newline
\noindent The Di\'{o}si-Penrose energy depends on the mass distributions of the superposition's states $\rho_{_{1}}(\mathbf{x})$ and $\rho_{_{2}}(\mathbf{x})$ as \cite{Pen-1,Dio-3,GExp-11}

\begin{equation}
\label{eq:2}
E_{_{G12}}=\xi G \int d^{3}\mathbf{x} d^{3}\mathbf{y}\frac{(\rho_{_{1}}(\mathbf{x})\mathsmaller{-}\rho_{_{2}}(\mathbf{x}))(\rho_{_{1}}(\mathbf{y})\mathsmaller{-}\rho_{_{2}}(\mathbf{y}))}{|\mathbf{x}-\mathbf{y}|}
 \textrm{\textsf{~~,~~~~~~\footnotesize \textit{Di\'{o}si-Penrose energy}}}
\end{equation}

~
\newline
\noindent where $G$ is the gravitational constant and $\xi$ a dimensionless parameter in the order of one. In their original publications, Di\'{o}si and Penrose derived this dimensionless parameter as $\xi$$=$$1$ \cite{Dio-3,Pen-1}. In an overview article, Bassi however showed that Di\'{o}sis approach leads to a Di\'{o}si-Penrose energy with $\xi$$=$$\frac{1}{2}$ \cite{Dio-5}. Here we will show that $\xi$ can be consistently derived from Di\'{o}si's and Penrose's approach and also from semiclassical gravity to be $\xi$$=$$\frac{1}{2}$.

\bigskip
\bigskip
\noindent
The Di\'{o}si-Penrose energy has different physical illustrations. One, which directly follows from Equation (\ref{eq:2}), is that it describes the gravitational self-energy resulting from the difference of the states' mass distributions {\small$\rho_{_{1}}(\mathbf{x})$$-$$\rho_{_{2}}(\mathbf{x})$}. This illustration is not very intuitive, since this difference can be negative. Despite this fact, the Di\'{o}si-Penrose energy is always positive, as we will see later. 

\bigskip
\noindent
A more helpful illustration of the Di\'{o}si-Penrose energy is as follows, which only holds for superposed rigid bodies, whose states are displaced against each other by a distance $\Delta s$ (i.e. {\small$\rho_{_{2}}(\mathbf{x})$$=$$\rho_{_{1}}(\mathbf{x}$$-$$\Delta s)$}). Assuming hypothetically that the masses of the superposition's states attract each other by the gravitational force, the Di\'{o}si-Penrose energy describes the mechanical work to pull the masses apart from each other over the distance of $\Delta s$ against their gravitational attraction. From this illustration, it follows that for small displacements $\Delta s$, where the gravitational force can be linearised, the Di\'{o}si-Penrose energy increases quadratically with the displacement $\Delta s$ ($E_{_{G12}}$$\propto$$\Delta s^{2}$). At large displacements $\Delta s$, where the gravitational attraction vanishes, the Di\'{o}si-Penrose energy converges to a constant value. It is important to note that this illustration of the Di\'{o}si-Penrose energy leads also to $\xi$$=$$\frac{1}{2}$ in Equation (\ref{eq:2}). The derivation of this illustration of the Di\'{o}si-Penrose energy is given in the appendix of \cite{Solid}.

\bigskip
\bigskip
\noindent
Sometimes, it is helpful to convert the Di\'{o}si-Penrose energy (Equation \ref{eq:2}) into a different form. With the gravitational potentials $\Phi_{_{i}}(\mathbf{x})$ resulting from the states' mass distributions $\rho_{_{i}}(\mathbf{x})$, which are given by

\begin{equation}
\label{eq:3}
\Phi_{_{i}}(\mathbf{x})=-G\int d^{3}\mathbf{y}\frac{\rho_{_{i}}(\mathbf{y})}{|\mathbf{x}-\mathbf{y}|}
\textrm{\textsf{~~,~~~~~~~~~~~~~~~~~}}
\end{equation}

~
\newline
\noindent Equation (\ref{eq:2}) can be converted for $\xi$$=$$\frac{1}{2}$ to

\begin{equation}
\label{eq:4}
E_{_{G12}}=\frac{1}{2} \int  d^{3}\mathbf{x}(\rho_{_{1}}(\mathbf{x})\mathsmaller{-}\rho_{_{2}}(\mathbf{x}))(\Phi_{_{2}}(\mathbf{x})\mathsmaller{-}\Phi_{_{1}}(\mathbf{x}))
\textrm{\textsf{~~,~~~~~~~\footnotesize \textit{Di\'{o}si-Penrose energy}}}
\end{equation}

~
\newline
\noindent i.e. the Di\'{o}si-Penrose energy is described by the integral of the difference of the states' mass distributions multiplied by the difference of their gravitational potentials.

\bigskip
%
%
%
\subsection{Penrose's approach }                 
%
%
\label{sec:2.2}
Penrose's approach \cite{Pen-1} is based on the argument that superposed states prefer (according to general relativity) differently curved spacetimes when their mass distributions are different, which leads to an uncertainty in the location in spacetime. From this uncertainty follows a fuzziness of the states' energies, which Penrose accounts for the superposition's decay.

\bigskip
\noindent
Before coming to Penrose's original derivation, an alternative derivation will be proposed, which is not as precise as Penrose's one, but expresses the idea quite well. In the Newtonian limit, only the $g_{_{00}}$-component of the metric field $g_{_{\mu \nu}}(x)$ is of relevance, which describes the derivation of the physical time according to the time coordinate $x^{0}$ as {\small$\frac{ds}{dx^{0}}$$=$$\sqrt{g_{_{00}}}$}. The $g_{_{00}}$-component can be expressed by the gravitational potential $\Phi (\mathbf{x})$ as follows \cite{Gen-7}:

\begin{equation}
\label{eq:5}
\frac{ds}{dx^{0}}=\sqrt{g_{_{00}}}\approx1+\frac{\Phi(\mathbf{x})}{c^{2}}
\textrm{\textsf{~~.~~~~~~~~~~~~}}
\end{equation}

~
\newline
\noindent Since superposed states with different mass distributions have different gravitational potentials, a clock runs with slightly different speeds depending on to whose state's spacetime geometry the clock is assigned. This uncertainty of the clock's speed leads to a fuzziness of the states' energies. Multiplying the difference of the clock's speed {\small$(\frac{ds_{1}}{dx^{0}}$$-$$\frac{ds_{2}}{dx^{0}})$} in State 1's and 2's  spacetime geometries by the energy density of State 1 $\rho_{_{1}}(\mathbf{x})c^{2}$, we obtain an estimate for the fuzziness of State 1's energy as follows:

\begin{equation}
\label{eq:6}
\Delta E_{_{1}}=\int d^{3}\mathbf{x}(\frac{ds_{_{2}}}{dx^{0}}-\frac{ds_{_{1}}}{dx^{0}})\rho_{_{1}}(\mathbf{x})c^{2}
 \textrm{~~}
=\int d^{3}\mathbf{x}\rho_{_{1}}(\mathbf{x})(\Phi_{_{2}}(\mathbf{x})-\Phi_{_{1}}(\mathbf{x}))
\textrm{\textsf{~~.}}
\end{equation}

~
\newline
\noindent In the same way, the fuzziness of state 2's energy yields {\small$\Delta E_{_{2}}$$=$$\int d^{3}\mathbf{x}\rho_{_{2}}(\mathbf{x})(\phi{_{1}}(\mathbf{x})$$-$$\phi{_{2}}(\mathbf{x}))$}. If we assume that in {\small$50\%$} of the cases State 1 decays to State 2, and in the other {\small$50\%$} of the cases vice versa, the relevant fuzziness of energy is {\small$\Delta E$$=$$(\Delta E_{_{1}}+\Delta E_{_{2}})/2$}, which is the Di\'{o}si-Penrose energy in the form of Equation (\ref{eq:4}). 

\bigskip
\noindent
Now we turn to Penrose's original derivation \cite{Pen-1}. Penrose compares the gravitational fields $\mathbf{g}_{_{i}}(\mathbf{x})$ that arise in the differently curved spacetimes of States 1 and 2 as follows:

\begin{equation}
\label{eq:7}
E_{_{G12}}=\frac{1}{8\pi G}\int d^{3}\mathbf{x}|\mathbf{g}_{_{1}}(\mathbf{x})-\mathbf{g}_{_{2}}(\mathbf{x})|^{2}
\textrm{\textsf{~~.~~~~~~~~~~~~}}
\end{equation}

~
\newline
\noindent The fuzziness of energy following from this approach can be physically justified with the energy density that one can assign the gravitational field, which is given by {\small$\frac{1}{8\pi G}\mathbf{g}^{2}(\mathbf{x})$}. With {\small$\mathbf{g}_{_{i}}(\mathbf{x})$$=$$-$$\nabla \Phi_{_{i}}(\mathbf{x})$} and converting Equation (\ref{eq:7}) with Green's first identity and Poisson's equation ({\small$\Delta \Phi(\mathbf{x})$$=$$4\pi G \rho(\mathbf{x})$}), we again obtain the Di\'{o}si-Penrose energy in the form of Equation (\ref{eq:4}). Penrose does not, in his derivation, refer to the energy density that one can assign the gravitational field ({\small$\frac{1}{8\pi G}\mathbf{g}^{2}(\mathbf{x})$}), and uses in Equation (\ref{eq:7})  the factor $\frac{1}{4\pi}$ instead of $\frac{1}{8\pi}$, which leads to a Di\'{o}si-Penrose energy with $\xi$$=$$1$ instead of $\xi$$=$$\frac{1}{2}$ in Equation (\ref{eq:2}).

\bigskip
\noindent
From Equation (\ref{eq:7}), it follows that the Di\'{o}si-Penrose energy is always positive. This is important, since otherwise the Di\'{o}si-Penrose criterion (Equation \ref{eq:1}) leads to negative lifetimes.

\bigskip
\bigskip
\noindent
A problematic point in Penrose's approach, which was addressed by Penrose himself, is as follows. To calculate the uncertainty of the gravitational field according to Equation (\ref{eq:7}), the points of the spacetime geometry of State 1 have to be identified with points of the spacetime geometry of State 2 for comparing the gravitational fields. Penrose sees in this procedure a fundamental problem, which he expresses in his own words as follows \cite{Pen-2}:

\begin{quote}
\textit{\small
The principle of general covariance tells us not only that there are to be no preferred coordinates, but also that, if we have two different spacetimes, representing two physically distinct gravitational fields, then there is to be no naturally preferred pointwise identification between the two - so we cannot say which particular spacetime point of one is to be regarded as the same point as some particular spacetime point of the other! 
}
\end{quote}

\bigskip
%
%
%
\subsection{Di\'{o}si's approach}                 
%
%
\label{sec:2.3}
Di\'{o}si adapts in his approach the work of Bohr and Rosenfeld, who investigated the uncertainty of measuring an electromagnetic field by an apparatus obeying quantum mechanics, to the measurement of a gravitational field. He found that a gravitational field measured over a time $\Delta t$ on a volume $V$ exhibits the following uncertainty $\delta g$ \cite{Dio-6}:

\begin{equation}
\label{eq:8}
(\delta g)^{2}\geq \frac{\hbar G}{V\Delta t}
\textrm{\textsf{~~.~~~~~~~~~~~~~~~~~~~~~}}
\end{equation}

~
\newline
\noindent Di\'{o}si postulates that the gravitational field exhibits universal fluctuations according to this uncertainty, i.e. a universal gravitational white noise. For the mathematical formulation of his approach, he uses the framework of the dynamical reduction models, which provide evolution equations for the density matrix, when stochastically fluctuating operators are introduced \cite{GRW_Ue-2}. This leads to the following equation of motion for the density matrix $\rho$ \cite{Dio-1,Dio-2}:

\begin{equation}
\label{eq:9}
\frac{d\rho}{dt}=-\frac{i}{\hbar}[H,\rho]-\frac{G}{2\hbar}\int \int d^{3}\mathbf{x}d^{3}\mathbf{y} \frac{1}{|\mathbf{x}-\mathbf{y}|}
[\hat{\rho}(\mathbf{x}),[\hat{\rho}(\mathbf{y}),\rho]]
\textrm{\textsf{~~,~~~~~~~~~~~}}
\end{equation}

~
\newline
\noindent in which $H$ is the Hamiltonian, and $\hat{\rho} (\mathbf{x})$ the operator of mass density. From this equation, one can derive a characteristic decay time $\tau_{_{d}}$ describing of how fast the interference between states of different mass distributions destroys. This decay time is given by the Diósi–Penrose criterion, i.e. by $\tau_{_{d}}$$=$$\hbar /E_{_{G12}}$, with a Di\'{o}si-Penrose energy $E_{_{G12}}$ (Equation \ref{eq:2}) with $\xi$$=$$\frac{1}{2}$ \cite{Dio-5}
\footnote{   
Di\'{o}si obtains in his original work a Di\'{o}si-Penrose energy with $\xi$$=$$1$ \cite{Dio-3}. In an overview article \cite{Dio-5}, Bassi however showed that Di\'{o}si's approach leads to a Di\'{o}si-Penrose energy with $\xi$$=$$\frac{1}{2}$, which is also intuitively expected by comparing Equations (\ref{eq:9}) and (\ref{eq:2}).
}.

\bigskip
\noindent
A problematic point in Di\'{o}si's approach is that the mass density operator $\hat{\rho} (\mathbf{x})$ in Equation (\ref{eq:9}) has to be modified to avoid divergences. The delta-shaped mass density operator has to be smeared, e.g. as in \cite{Dio-1}

\begin{equation}
\label{eq:10}
\hat{\rho}(\mathbf{x})=\sum_{i}m_{_{i}}\delta(\mathbf{x}-\hat{\mathbf{x}}_{_{i}})
\textrm{~~~}
\Rightarrow
\textrm{~~~}
\hat{\rho}(\mathbf{x})=\sum_{i}\frac{m_{_{i}}}{\frac{4\pi}{3}r^{3}_{0}}\Theta(r_{_{0}}-|\mathbf{x}-\hat{\mathbf{x}}_{_{i}}|)
\textrm{\textsf{~~,~~~~~~~}}
\end{equation}

~
\newline
\noindent where $r_{_{0}}$ is the characteristic radius for the smearing and $m_{_{i}}$, $\hat{\mathbf{x}}_{_{i}}$ the mass and position operator of the i's particle. The characteristic radius $r_{_{0}}$ was originally chosen by Di\'{o}si to be on the order of the nucleon's radius, i.e. $r_{_{0}}$$\approx$$10^{-13}cm$ \cite{Dio-1}, and was later revised by Ghirardi to a much larger value of $r_{_{0}}$$\approx$$10^{-15}cm$ \cite{Dio-2}. This correction was necessary to avoid a too-strong permanent increase of total energy \cite{Dio-2}. A permanent increase of total energy is characteristic of dynamical reduction models, and results from stochastically fluctuating operators \cite{GRW_Ue-2}.

\newpage
%
%
%
\section{Semiclassical gravity}                 
%
%
\label{sec:3}
We begin our analysis of semiclassical gravity by regarding first two-state superpositions in the Newtonian limit. In Section \ref{sec:3.1}, we calculate the total energy of such superpositions. In Section \ref{sec:3.2}, we show how the states' energies increase due to the sharing of spacetime in semiclassical gravity. In Section \ref{sec:3.3}, we derive the Di\'{o}si-Penrose criterion and Born's rule from this result.

%
%
%
\ \subsection{Total energy of a two-state superposition}                 
%
%
\label{sec:3.1}
The state vector $|\psi$$>$ of the single-photon experiment in Figure \ref{fig1} can be thought of as a localised wavepacket in configuration space, which splits into two well separated wavepackets $|\psi_{_{1}}$$>$ and $|\psi_{_{2}}$$>$ when the photon enters the beam splitter at $t_{_{s}}$, as shown in the right-hand side of Figure \ref{fig1}. The state vector $|\psi$$>$ will hereafter describe the entire system, consisting for the experiment in Figure \ref{fig1} of the photon, the beam splitter, the detector and the rigid body the detector is displacing. After the photon was split by the beam splitter, the state vector can be written as a superposition of the wavepackets $|\psi_{_{1}}$$>$ and $|\psi_{_{2}}$$>$ as

\begin{equation}
\label{eq:11}
|\psi>=c_{_{1}}|\psi_{_{1}}>+c_{_{2}}|\psi_{_{2}}>
\textrm{\textsf{~~,~~~~~~~~~~~~~~~~~~~}}
\end{equation}

~
\newline
\noindent where $|\psi_{_{1}}$$>$ and $|\psi_{_{2}}$$>$ correspond to the cases that the photon was reflected and respectively transmitted at the beam splitter. The amplitudes $c_{_{1}}$ and $c_{_{2}}$ of this superposition fulfil the following normalisation:

\begin{equation}
\label{eq:12}
|c_{_{1}}|^{2}+|c_{_{2}}|^{2}=1
\textrm{\textsf{~~,~~~~~~~~~~~~~~~~~~~~~~~~}}
\end{equation}

~
\newline
\noindent since $|\psi_{_{1}}$$>$ and $|\psi_{_{2}}$$>$ shall be normalised ($<$$\psi_{_{i}}|\psi_{_{i}}$$>$$=$$1$). The mass distributions $\rho_{_{i}}(\mathbf{x})$ of the states $|\psi_{_{1}}$$>$ and $|\psi_{_{2}}$$>$ can be calculated with the operator of mass density $\hat{\rho} (\mathbf{x})$ as follows:

\begin{equation}
\label{eq:13}
\rho_{_{i}}(\mathbf{x})=<\psi_{_{i}}|\hat{\rho}(\mathbf{x})|\psi_{_{i}}>
\textrm{\textsf{~~.~~~~~~~~~~~~~~~~~~~~~~~~}}
\end{equation}

~
\newline
\noindent When the wavepackets corresponding to $|\psi_{_{1}}$$>$ and $|\psi_{_{2}}$$>$ are well separated in configuration space, as in the right-hand side of Figure \ref{fig1} for $t\,$$>$$ t_{_{s}}$, the mass distribution of the superposition is given by the mean of the states' mass distributions as follows.  

\begin{equation}
\label{eq:14}
\rho(\mathbf{x})=|c_{_{1}}|^{2}\rho_{_{1}}(\mathbf{x})+|c_{_{2}}|^{2}\rho_{_{2}}(\mathbf{x})
\textrm{\textsf{~~.~~~~~~~~~~~~~~~~~~~}}
\end{equation}

~
\newline
\noindent Since spacetime geometry is treated classically in semiclassical gravity, and the metric field is in the Newtonian limit directly linked to the gravitational potential (cf. Equation \ref{eq:5}), the gravitational potential must also be treated classically. The gravitational potential resulting from the mass distribution of Equation (\ref{eq:14}) is:

\begin{equation}
\label{eq:15}
\Phi(\mathbf{x})=|c_{_{1}}|^{2}\Phi_{_{1}}(\mathbf{x})+|c_{_{2}}|^{2}\Phi_{_{2}}(\mathbf{x})
\textrm{\textsf{~~,~~~~~~~~~~~~~~~~~~~}}
\end{equation}

~
\newline
\noindent where $\Phi_{_{1}}(\mathbf{x})$ and $\Phi_{_{2}}(\mathbf{x})$ are the gravitational potentials resulting from the states' mass distributions according to Equation (\ref{eq:3}). This means that the gravitational potential of the superposition is given by the mean of the states' gravitational potentials. 

\bigskip
\noindent
The total energy of our two-state superposition can be calculated with

\begin{equation}
\label{eq:16}
E=\int d^{3}\mathbf{x}\rho (\mathbf{x})(c^{2}+\mathsmaller{\frac{1}{2}}\Phi (\mathbf{x}))
\textrm{\textsf{~~,~~~~~~~~~~~~~~~~~~~}}
\end{equation}

~
\newline
\noindent where $\rho (\mathbf{x})$ and $\Phi (\mathbf{x})$ are the mass distribution and gravitational potential, respectively, according to Equations (\ref{eq:14}) and (\ref{eq:15}). The first term in Equation (\ref{eq:16}) calculates the masses' rest energies ($E$$=$$mc^{2}$). The masses' kinetic energies are neglected, since they are small compared to the rest energies. The second term calculates the gravitational energies between the masses ($E_{_{G}}$$=$$-$$Gm_{_{1}}m_{_{2}}/r$), where the factor $\frac{1}{2}$ avoids gravitational energy between the two masses being counted twice during integration. By inserting Equations (\ref{eq:14}) and (\ref{eq:15}) into Equation (\ref{eq:16}), we obtain after a short calculation
\footnote{   
The second term in Equation (\ref{eq:16}) can be transformed with the normalisation $|c_{_{1}}|^{2}+|c_{_{2}}|^{2}=1$ as follows:\\ 
\noindent {\small$( |c_{_{1}}|^{2} \rho_{_{1}} + |c_{_{2}}|^{2} \rho_{_{2}}) \dot (|c_{_{1}}|^{2} \Phi_{_{1}} + |c_{_{2}}|^{2} \Phi_{_{2}}) = \\
|c_{_{1}}|^{2} \rho_{_{1}} (|c_{_{1}}|^{2} \Phi_{_{1}} + |c_{_{2}}|^{2} \Phi_{_{2}}) + |c_{_{2}}|^{2} \rho_{_{2}} (|c_{_{1}}|^{2} \Phi_{_{1}} + |c_{_{2}}|^{2} \Phi_{_{2}}) = \\
|c_{_{1}}|^{2} \rho_{_{1}} (\Phi_{_{1}} - |c_{_{2}}|^{2} \Phi_{_{1}} + |c_{_{2}}|^{2} \Phi_{_{2}}) + |c_{_{2}}|^{2} \rho_{_{2}} (|c_{_{1}}|^{2} \Phi_{_{1}} + \Phi_{_{2}} - |c_{_{1}}|^{2} \Phi_{_{2}}) = \\
|c_{_{1}}|^{2} \rho_{_{1}}\Phi_{_{1}} + |c_{_{2}}|^{2} \rho_{_{2}}\Phi_{_{2}} + |c_{_{1}}|^{2} |c_{_{2}}|^{2} (\rho_{_{1}} - \rho_{_{1}}) (\Phi_{_{2}} - \Phi_{_{1}})$}.
}
the following total energy of our two-state superposition:

\begin{equation}
\label{eq:17}
E=|c_{_{1}}|^{2}E_{_{1}}+|c_{_{2}}|^{2}E_{_{1}}+|c_{_{1}}|^{2}|c_{_{2}}|^{2}E_{_{G12}}
\textrm{\textsf{~~.~~~~~~~~~~~~~~~~}}
\end{equation}

~
\newline
\noindent Here $E_{_{1}}$ and $E_{_{2}}$ are the total energies of States 1 and 2 alone, which, similar to Equation (\ref{eq:16}), are given by

\begin{equation}
\label{eq:18}
E_{i}=\int d^{3}\mathbf{x}\rho_{_{i}} (\mathbf{x})(c^{2}+\mathsmaller{\frac{1}{2}}\Phi_{_{i}} (\mathbf{x}))
\textrm{\textsf{~~.~~~~~~~~~~~~~~~~~~~~~~}}
\end{equation}

~
\newline
\noindent The term $|c_{_{1}}|^{2}|c_{_{2}}|^{2}E_{_{G12}}$, in which $E_{_{G12}}$ is the Di\'{o}si-Penrose energy according to Equation (\ref{eq:4}), expresses how much the superposition's total energy increases due to the sharing of the mean gravitational potential by the states. Since the sharing of gravitational potential expresses the sharing of spacetime geometry in semiclassical gravity, the Di\'{o}si-Penrose energy can be regarded as a measure of how much the preferred spacetime geometries of States 1 and 2 differ from each other, or of how strong the states compete for spacetime geometry.

\bigskip
\noindent
Since the Di\'{o}si-Penrose energy in Equation (\ref{eq:4}) corresponds to a factor of $\xi$$=$$\frac{1}{2}$ in Equation (\ref{eq:2}), semiclassical gravity leads as the approaches of Penrose and Di\'{o}si to a Di\'{o}si-Penrose energy with $\xi$$=$$\frac{1}{2}$.

\newpage
%
%
%
\subsection{Energy increases of the states}                 
%
%
\label{sec:3.2}
In the previous section, we investigated how the sharing of spacetime geometry in semiclassical gravity increases the total energy of a superposition. In this section, we investigate this from the point of view of the states of the superposition. We calculate how much the energies of States 1 and 2 of our two-state superposition increase due to the sharing of a common gravitational potential in semiclassical gravity.

\bigskip
\noindent
The total energy of e.g. State 1 is given by the terms in the superposition's total energy (Equation \ref{eq:17}), which are proportional to its intensity $|c_{_{1}}|^{2}$, which are $|c_{_{1}}|^{2}E_{_{1}}$ and $|c_{_{1}}|^{2}|c_{_{2}}|^{2}E_{_{G12}}$, and by normalising these terms to this intensity, which yields a total energy of $E_{_{1}}$$+$$|c_{_{2}}|^{2}E_{_{G12}}$. In the same way, the total energy of State 2 is $E_{_{2}}$$+$$|c_{_{1}}|^{2}E_{_{G12}}$. The energy increases of States 1 and 2 with respect to the case that they are alone and must not share spacetime geometry  (gravitational potential) with the other are given by

\begin{equation}
\label{eq:19}
\begin{split}
E_{_{G1}}=|c_{_{2}}|^{2}E_{_{G12}} \\
E_{_{G2}}=|c_{_{1}}|^{2}E_{_{G12}}  
\end{split}
\textrm{\textsf{~~~~.~~~~~~~~~~~~~~\footnotesize \textit{energy increases of states}}}
\end{equation}

\begin{figure}[h]
\centering
\includegraphics[width=9cm]{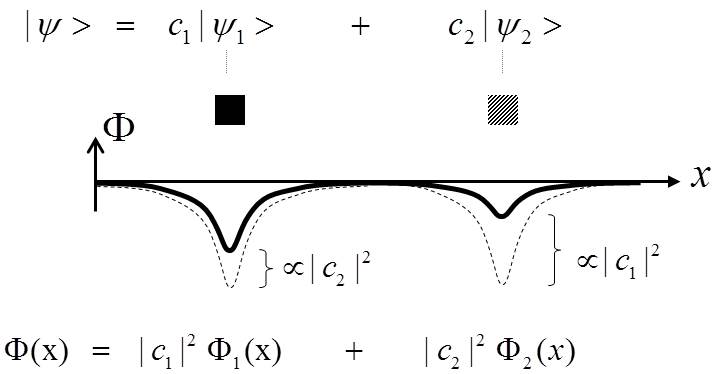}\vspace{0cm}
\caption{\footnotesize
\mbox{Gravitational potential of the superposition generated by the single-photon experiment in Figure \ref{fig1}.}
}
\label{fig2}
\end{figure}

~
\newline
\noindent This means that the energy increases of the states are proportional to the intensity of the respective competing state. This can be seen in Figure \ref{fig2}, illustrating the mean gravitational potential of the superposition generated by the single-photon experiment in Figure \ref{fig1}. Each state suffers due to the sharing of the mean gravitational potential with the other state an increase of its gravitational potential, proportional to the intensity of the respective competing state.

%

\bigskip
\noindent
From Figure \ref{fig2}, one can also see that the energy increases (Equation \ref{eq:19}) are proportional to the Di\'{o}si-Penrose energy $E_{_{G12}}$. For a large displacement between the states, the Di\'{o}si-Penrose energy (Equation \ref{eq:4}) can be written as {\small$E_{_{G12}}$$=$$-$$\int  d^{3}\mathbf{x}\rho_{_{1}}(\mathbf{x})\Phi_{_{1}}(\mathbf{x})$}, respective {\small$E_{_{G12}}$$=$$-$$\int  d^{3}\mathbf{x}\rho_{_{2}}(\mathbf{x})\Phi_{_{2}}(\mathbf{x})$}.

\newpage
%
%
%
\subsection{Di\'{o}si-Penrose criterion and Born's rule}                 
%
%
\label{sec:3.3}
In this section, we show how one can establish a connection between the energy increases of the states resulting from semiclassical gravity (Equation \ref{eq:19}) and the Di\'{o}si-Penrose criterion and Born's rule. For this, we hypothetically assume that the states' energy increases $E_{_{Gi}}$ are responsible for their decay. We define for every state from its energy increase $E_{_{Gi}}$ a decay rate as follows:

\begin{equation}
\label{eq:20}
\begin{split}
\frac{dp_{_{1\downarrow}}}{dt}&=\frac{E_{_{G1}}}{\hbar} \\
\frac{dp_{_{2\downarrow}}}{dt}&=\frac{E_{_{G2}}}{\hbar}
\end{split}
\textrm{\textsf{~~~~,~~~~~~~~~~~~~~\footnotesize \textit{decay rates of states}}}
\end{equation}

 ~
\newline
\noindent where $dp_{_{i\downarrow}}$ is the probability for a decay of state $i$ during the time interval $dt$. With Equation (\ref{eq:19}), the decay rates can be converted to

\begin{equation}
\label{eq:21}
\begin{split}
\frac{dp_{_{1\downarrow}}}{dt}&=|c_{_{2}}|^{2}\frac{E_{_{G12}}}{\hbar} \\
\frac{dp_{_{2\downarrow}}}{dt}&=|c_{_{1}}|^{2}\frac{E_{_{G12}}}{\hbar}
\end{split}
\textrm{\textsf{~~~~.~~~~~~~~~~~~}}
\end{equation}

\bigskip   
\bigskip
\bigskip
\noindent
\textbf{Di\'{o}si-Penrose criterion} \\  
\noindent
The mean lifetime $T_{_{G}}$ of our two-state superposition can be calculated with the decay rates (Equation \ref{eq:21}) and the normalisation $|c_{_{1}}|^{2}+|c_{_{2}}|^{2}=1$ as follows:

\begin{equation}
\label{eq:22}
\frac{1}{T_{_{G}}}=\frac{dp_{_{1\downarrow}}}{dt}+\frac{dp_{_{2\downarrow}}}{dt}=\frac{E_{_{G12}}}{\hbar}
\textrm{\textsf{~~,~~~~~~~~~~~~~~\footnotesize \textit{Di\'{o}si-Penrose criterion}}}
\end{equation}

~
\newline
\noindent which complies with the Di\'{o}si-Penrose criterion (Equation \ref{eq:1}). That there is a relationship between semiclassical gravity and the Di\'{o}si-Penrose criterion was recently shown by Bera et al. with a different kind of analysis based on a stochastic modification of the Schr\"odinger-Newton equation \cite{NS-16}.

\bigskip   
\bigskip
\bigskip
\noindent
\textbf{Born's rule} \\  
\noindent
Since a decay of State 1 leads to a reduction to State 2 and vice versa, and since the decay rates are, according to Equation (\ref{eq:21}), proportional to the intensity of the competing state in favour of which the state decays, we obtain reduction probabilities proportional to the states' intensities, i.e. Born's rule:

\begin{equation}
\label{eq:23}
\begin{split}
p_{_{1}}=|c_{_{1}}|^{2} \\
p_{_{2}}=|c_{_{2}}|^{2}
\end{split}
\textrm{\textsf{~~.~~~~~~~~~~~~~~\footnotesize \textit{Born's rule}}}
\end{equation}

~
\newline
\noindent This means that Born's rule could have its physical origin in semiclassical gravity, in the way that the energy increase, respectively the decay rate, of a state resulting from the sharing of spacetime geometry (respectively gravitational potential) in semiclassical gravity is proportional to the intensity of the competing state in favour of which the state decays. This supposition will be followed up in Section \ref{sec:5}.

\bigskip
%
%
%
\subsection{Perspectives for semiclassical gravity}                 
%
%
\label{sec:3.4}
In this section, we show that semiclassical gravity, on which the Dynamical Spacetime approach to wavefunction collapse derived in Part 2 is based on, is a solid basis for a reduction model, since it does not lead to problems occurring in the approaches of Penrose and Di\'{o}si.

\bigskip
\noindent
The problematic point in Penrose's approach to identify points of different spacetimes with each other (cf. end of Section \ref{sec:2.2}) plays no role in semiclassical gravity, since one has only one spacetime geometry. 

\bigskip
\noindent
There is also no need to modify the operator of mass density, as in Di\'{o}si's approach (cf. Equation \ref{eq:10}). The states' mass distributions needed for the calculation of the Di\'{o}si-Penrose energy can be determined as in Penrose's approach with the unmodified delta-shaped mass density operator (Equation \ref{eq:10}) without divergences \cite{Dio-5}. Concrete examples for calculating Di\'{o}si-Penrose energies in realistic contexts are given in \cite{Solid}, in which the Di\'{o}si-Penrose energies of solids in quantum superposition are calculated, where the calculations take the solid's microscopic mass distribution resulting from its nuclei into account.

\newpage
%
%
%
\section{Relativistic generalisation}                 
%
%
\label{sec:4}
In this section, we show how the results of Section \ref{sec:3} can be relativistically generalised. We are looking for the relativistic generalisation of the Di\'{o}si-Penrose energy and of the states' decay rates, which led us to the Di\'{o}si-Penrose criterion and Born's rule. Since energy depends on the chosen Lorentz frame, the Di\'{o}si-Penrose energy itself is not a suitable quantity for relativistic generalisation. Therefore, we construct from the Di\'{o}si-Penrose energy a so-called \textit{competition action} measuring the difference between the states' preferred spacetime geometries on a spacetime region instead of only at a point in time, such as the Di\'{o}si-Penrose energy. We show that the relativistic generalisation of this competition action arises from a decomposition of the Einstein-Hilbert action according to so-called \textit{classical scenarios}.

\bigskip
\noindent
This section is structured as follows. In Section \ref{sec:4.1}, we recapitulate the basics of relativistic quantum mechanics; in particular, the Tomonaga-Schwinger equation. In Section \ref{sec:4.2}, we define the classical scenarios resembling approximately classical trajectories of the system, and according to which the state vector's evolution can be decomposed. In Section \ref{sec:4.3}, we introduce the competition action, and show how the states' decay rates can be expressed with this action. In Section \ref{sec:4.4}, we decompose the Einstein-Hilbert action according to the classical scenarios and derive the relativistic generalisation of the competition action. In Section \ref{sec:4.5}, we derive the relativistic generalisation of the states' decay rates.

\bigskip
\bigskip
%
%
%
\subsection{Tomonaga-Schwinger equation }                 
\begin{figure}[t]
\centering
\includegraphics[width=12cm]{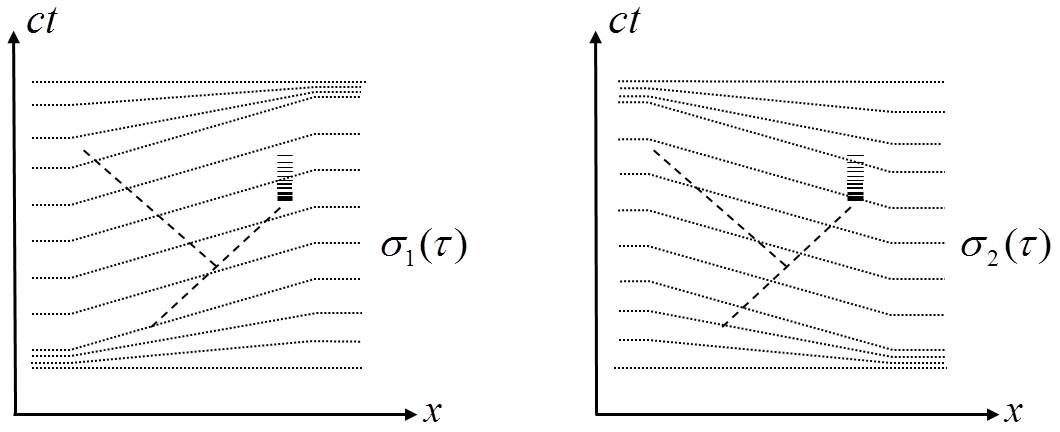}\vspace{0cm}
\caption{\footnotesize
Two hypersurface sequences for the single-photon experiment in Figure \ref{fig1}. Illustrated are the photon line and the detection process in the detector.
}
\label{fig3}
\end{figure}

%
%
\label{sec:4.1}
The Tomonaga-Schwinger was developed in the context of quantum field theory \cite{Gen-1,Gen-2}. It can be understood as the relativistic equivalent of Schr\"odinger's equation. With the Tomonaga-Schwinger equation, the evolution of state vector describing the quantum fields can be followed up on arbitrarily chosen sequences of spacelike hypersurfaces $\sigma (\tau )$, where $\tau$ is a dimensionless parameter parametrising the sequence \cite{Gen-1,Gen-2}. Figure \ref{fig3} illustrates two such hypersurface sequences ($\sigma_{_{1}}(\tau )$, $\sigma_{_{2}}(\tau )$) for the single-photon experiment in Figure \ref{fig1}, where the figure shows the photon line and the detection process in the detector. Here we restrict ourselves to time-ordered hypersurface sequences, which means that all points of $\sigma (\tau $$+$$d\tau )$ are in the future with respect to $\sigma (\tau )$. The evolution of the state vector $|\psi (\tau )$$>$ on $\sigma (\tau )$ is given by the Tomonaga-Schwinger equation as \cite{Gen-1,Gen-2,GRW_relW-7}

\begin{equation}
\label{eq:24}
|\psi (\tau +d\tau )>=T[e^{\mathlarger{-\frac{i}{\hbar}\int^{\sigma (\tau+d\tau)}_{\sigma(\tau)}d^{4}x H(x)}}] |\psi (\tau)>
\textrm{\textsf{~~,~~~~~~~~~~~}}
\end{equation}

~
\newline
\noindent where $H(x)$ is the Hamiltonian density, and $T[]$ an operator describing path ordering of timelike separated events
\footnote{   
I.e. the products of operators $...H(x_{_{i+1}})dx_{_{i+1}} \cdot H(x_{_{i}})dx_{_{i}} \cdot H(x_{_{i-1}})dx_{_{i-1}}...$ have to be ordered in a way that points $x_{_{i}}$ are never in the future light cones of subsequent points $x_{_{i+n}}$.
}.
The Lorentz invariance of the Tomonaga-Schwinger equation is reflected in the fact that results are independent of the chosen hypersurface sequence $\sigma (\tau )$. In the example of Figure \ref{fig3}, one obtains for the sequences $\sigma_{_{1}}(\tau )$ and $\sigma_{_{2}}(\tau )$, displaying Lorentz frames with positive and negative velocities on the spacetime region where the experiment is performed, the same final state vector. Physical quantities, such as e.g. the energy momentum tensor field $T_{_{\mu \nu}}(x)$, can be calculated with suitably defined operators as

\begin{equation}
\label{eq:25}
T_{_{\mu \nu}}(x)=<\psi (\sigma)|\hat{T}_{_{\mu \nu}}(x)|\psi(\sigma)>
\textrm{\textsf{\footnotesize~~,~~~~\textit{with}~~~~}}
x\in \sigma
\textrm{~,}
\end{equation}

~
\newline
\noindent where the result is independent of the chosen hypersurface $\sigma$.

%

\bigskip
%
%
%
\subsection{Classical scenarios }                 
%
%
\label{sec:4.2}
In this section, we decompose the state vector's evolution into evolutions resembling approximately classical trajectories of the system; the so-called \textit{classical scenarios}. We are looking for a decomposition of the state vector's evolution $|\psi (\tau )$$>$ as follows:

\begin{equation}
\label{eq:26}
|\psi(\tau)>=\sum_{i} c_{_{i}} |\tilde{\psi}_{_{i}} (\tau)>
\textrm{\textsf{\footnotesize~~,~~~~~~\textit{with}~~~~}}
\sum_{i} |c_{_{i}}|^{2}=1
\textrm{\textsf{~~,~~~~~~~~~~~}}
\end{equation}

\begin{figure}[h]
\centering
\includegraphics[width=16cm]{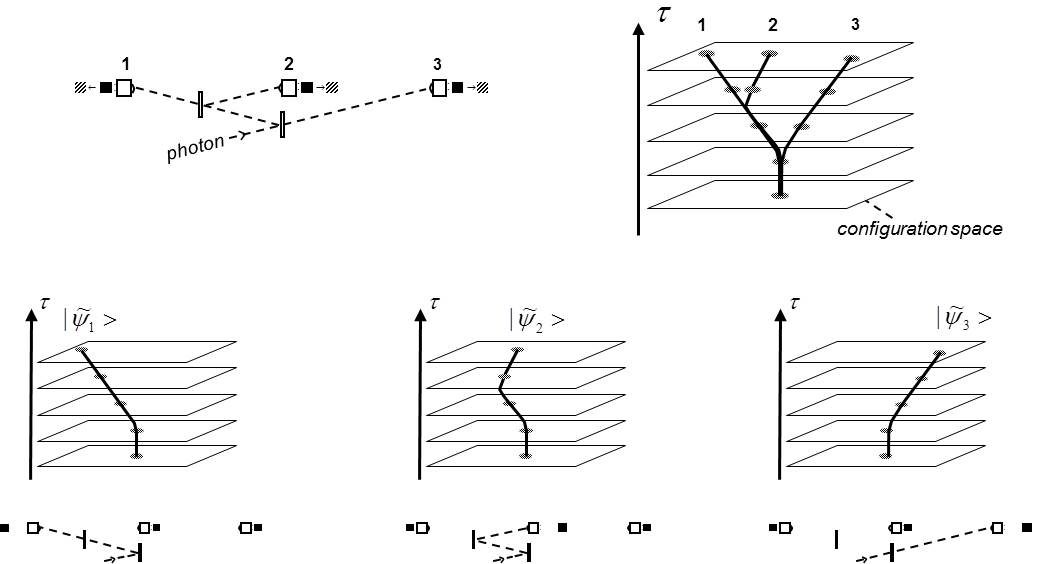}\vspace{0cm}
\caption{\footnotesize
Three classical scenarios $|\tilde{\psi}_{_{1}}(\tau )$$>$, $|\tilde{\psi}_{_{2}}(\tau )$$>$ and $|\tilde{\psi}_{_{3}}(\tau )$$>$ (middle part) of the three-detector experiment in the upper left, which are defined by following up the state vector's evolution $|\psi (\tau )$$>$ on classical paths in configuration space (upper right), for which the system evolves on classical trajectories, as shown at the bottom.
}
\label{fig4}
\end{figure}

~
\newline
\noindent where {\small$|\tilde{\psi}_{_{i}}(\tau )$$>$} are the classical scenarios that shall be normalised ({\small$<$$\tilde{\psi}_{_{i}}(\tau )|\tilde{\psi}_{_{i}}(\tau )$$>$$=$$1$}). The upper right part of Figure \ref{fig4} shows the state vector's evolution in configuration space for the three-detector experiment in the upper left of the figure, where the state vector's evolution is calculated with the Tomonaga-Schwinger equation for an arbitrarily chosen hypersurface sequence $\sigma (\tau )$ over the sequence parameter $\tau$. The lower part of the figure shows the evolutions of the three classical scenarios {\small$|\tilde{\psi}_{_{1}}(\tau )$$>$}, {\small$|\tilde{\psi}_{_{2}}(\tau )$$>$} and {\small$|\tilde{\psi}_{_{3}}(\tau )$$>$} in configuration space, and the corresponding behaviour in spacetime. The classical scenarios are defined by following up the state vector's evolution on classical paths in configuration space, for which the system evolves on classical trajectories in spacetime. In e.g. classic Classical Scenario 1 {\small$|\tilde{\psi}_{_{1}}(\tau )$$>$}, the photon is completely reflected at the first, and completely transmitted at the second beam splitter, and only detected by Detector 1, as illustrated in the lower left of Figure \ref{fig4}. The classical scenarios are not solutions of the Tomonaga-Schwinger equation with the parameters $\tau$, at which the state vector $|\psi (\tau )$$>$ splits into two wavepackets in configuration space, which is the case when the photon enters a beam splitter. To fulfil the decomposition of the state vector's evolution according to Equation (\ref{eq:26}) in the regions where several classical scenarios refer to the same root wavepacket, their phases must be chosen suitably. In e.g. the common root wavepacket of all classical scenarios, their phases must satisfy $|\sum_{i}|c_{_{i}}|e^{i\varphi_{i}}|$$=$$1$. Our convention of classical scenarios is independent of the chosen hypersurface sequence $\sigma (\tau )$, and can be regarded as Lorentz invariant. Classical scenarios play an important role in the formulation of the Dynamical Spacetime approach in Part 2 \cite{P2}.
\bigskip

%

\bigskip
\noindent
For the decomposition of the Einstein-Hilbert action according to classical scenarios in Section \ref{sec:4.4}, we will benefit from the following property of the classical scenario convention. It allows for the decomposition of a physical quantity, such as e.g. the energy momentum tensor field $T_{_{\mu \nu}}(x)$, in the entire spacetime region as follows:

\begin{equation}
\label{eq:27}
T_{_{\mu \nu}}(x)=\sum_{i} |c_{_{i}}|^{2} T_{_{\mu \nu \,i}}(x)
\textrm{\textsf{~~,~~~~~~~~~~~~~~~~~}}
\end{equation}

~
\newline
\noindent where $T_{_{\mu \nu i}}(x)$ are the classical scenarios' energy momentum tensor fields, calculated with the classical scenarios' state vectors {\small$|\tilde{\psi}_{_{i}}(\sigma )$$>$} according to  Equation (\ref{eq:25}).

\newpage
%
%
%
\subsection{Competition action }                 
%
%
\label{sec:4.3}
In this section, we define a measure of how much the preferred spacetime geometries of two classical scenarios differ from each other, which we call the \textit{competition action}. The competition action is defined by integrating the Di\'{o}si-Penrose energy $E_{_{G12}}(t)$ between the classical scenarios' states $|\psi_{_{1}}$$>$ and $|\psi_{_{2}}$$>$ over time as follows:

\begin{equation}
\label{eq:28}
\begin{split}
S_{_{G12}}(t)=\int^{t}_{..} dt E_{_{G12}}(t) ~~~~~~~~~~~~~~~~~~~~~~~~~~~~~~~~~~~~~~~~~~~~~~~~~~~~~~~~~~~~~~~~~~~~~~~~~~~~ \\
= \frac{1}{2} \int^{t}_{..} dt   \int  d^{3}\mathbf{x}(\rho_{_{1}}(\mathbf{x},t)-\rho_{_{2}}(\mathbf{x},t))(\Phi_{_{2}}(\mathbf{x},t)-\Phi_{_{1}}(\mathbf{x},t)) ~~.~~~~~~~~~~~ \\
 \textrm{\textsf{~~~~~~~~~~~~~~~~~~~~~~~~~~~~~~~~~~~~~~~~~~~~~~~\footnotesize \textit{competition action (Newtonian limit)}}}
\end{split}
\end{equation}

~
\newline
\noindent Furthermore, we define the "classical scenario"-equivalent of the states' energy increases $E_{_{Gi}}(t)$ by integrating them over time as follows:

\begin{equation}
\label{eq:29}
\begin{split}
S_{_{G1}}(t)\equiv \int^{t}_{..} dt E_{_{G1}}(t)\\
S_{_{G2}}(t)\equiv \int^{t}_{..} dt E_{_{G2}}(t) 
\end{split}
\textrm{\textsf{~~~~,~~~~~~~\footnotesize \textit{detuning actions (Newtonian limit)}}}
\end{equation}

~
\newline
\noindent which we call the \textit{detuning actions} of the classical scenarios. The detuning actions play an important role in the mathematical formulation of the Dynamical Spacetime approach \cite{P2}. The detuning actions of two classical scenarios can be expressed by the competition action between them as follows (cf. Equations \ref{eq:29} and \ref{eq:19}): 

\begin{equation}
\label{eq:30}
\begin{split}
S_{_{G1}}(t)=|c_{_{2}}|^{2}S_{_{G12}}(t) \\
S_{_{G2}}(t)=|c_{_{1}}|^{2}S_{_{G12}}(t) 
\end{split}
\textrm{\textsf{~~~~.~~~~~~~~~~~~~}}
\end{equation}

~
\newline
\noindent The classical scenarios' detuning actions are, like the states' energy increases (Equation \ref{eq:19}), proportional to the intensity of the respective competing scenario.

\bigskip
\noindent
With the classical scenarios' detuning actions $S_{_{Gi}}(t)$, the states' decay rates (Equation \ref{eq:20}) can be converted to

\begin{equation}
\label{eq:31}
\begin{split}
\frac{dp_{_{1\downarrow}}}{dt}=\frac{\frac{d}{dt}S_{_{G1}}(t)}{\hbar}  \\
\frac{dp_{_{2\downarrow}}}{dt}=\frac{\frac{d}{dt}S_{_{G2}}(t)}{\hbar} 
\end{split}
\textrm{\textsf{~~.~~~~~~~~~~~~~~\footnotesize \textit{decay rates of states (Newtonian limit)}}}
\end{equation}

~
\newline
\noindent This result has an intuitive physical illustration. The decay probability of a state $dp_{_{i\downarrow}}$ on a time interval $dt$ is given by the increase of its detuning action $dS_{_{Gi}}$ during this period of time divided by Planck's quantum of action $\hbar$. This result will be used as the starting point for the relativistic generalisation of the states' decay rates in Section \ref{sec:4.5}.

\newpage
%
%
%
\subsection{Relativistic generalisation of the competition action}                 
%
%
\label{sec:4.4}
In this section, we derive the relativistic generalisation of the competition action. For this, we decompose the Einstein-Hilbert action of a superposition according to our classical scenarios, which leads to an interaction term merging in the Newtonian into the competition action defined in the previous section.

\bigskip
\noindent
The Einstein-Hilbert action on a spacetime region, which is bounded towards the future by the hypersurface $\sigma (\tau )$, is given by \cite{Gen-7}:

\begin{equation}
\label{eq:32}
S_{_{EH}}(\tau)=\int^{\sigma (\tau)}_{...}\frac{d^{4}x}{c}\sqrt{-g(x)}\left(\frac{R(x)}{2\kappa} + L_{_{M}}(x)  \right)
\textrm{\textsf{~~,~~~~~~~~~~~~~~~~~}}
\end{equation}

~
\newline
\noindent where {\small  $d^{4}x\sqrt{-g(x)}$ }
\footnote{   
$g(x)$$\mathsmaller{\equiv}$$det(g_{_{\mu \nu}}(x))$.
}
is the covariant volume element, $R(x)$ the tension scalar, $L_{_{M}}(x)$ the Lagrangian density of all matter fields, and $\kappa=8\pi G/c^{4}$. The factor $1/c$ in the volume element is introduced to obtain the correct dimension of action (i.e. {\small$energy\cdot time$}).

\bigskip
\noindent
For weak gravitational fields for which Einstein's field equations can be linearised, and which are assumed here and in the Dynamical Spacetime approach in Part 2 \cite{P2}, the metric field $g_{_{\mu \nu}}(x)$ can be written as \cite{Gen-7}

\begin{equation}
\label{eq:33}
g_{_{\mu \nu}}(x)=\eta_{_{\mu \nu}}+h_{_{\mu \nu}} (x)
\textrm{\textsf{\footnotesize~~~~~~~with~~~~~~~}}
|h_{_{\mu \nu}}(x)|<<1
\textrm{~~,~~~~~~~~~~~~}
\end{equation}

~
\newline
\noindent where $\eta_{_{\mu \nu}}$ is the Minkowski metric ($\eta_{_{\mu \nu}}$$\mathsmaller{\equiv}$$diag(1,-1,-1,-1)$).

\bigskip
\noindent
For the calculation of the Einstein-Hilbert action, we have to determine the tension scalar $R(x)$ and the factor {\small$\sqrt{-g(x)}$}, which are both related to state vector's evolution $|\psi (\tau )$$>$ in the limit of weak gravitational fields as follows ($L$ indicates linear transformations):

\begin{equation}
\label{eq:34}
|\psi (\tau )> \Rightarrow T_{_{\mu \nu}}(x) \stackrel{L}{\Rightarrow}  h_{_{\mu \nu}}(x)
\begin{array} {c}
\stackrel{L}{\Rightarrow}R(x) ~~~~~~~~~~~  \\
\stackrel{L}{\Rightarrow} \sqrt{-g(x)} -1
\end{array}
\textrm{~~.~~~~~~~}
\end{equation}

~
\newline
\noindent The transformation from the state vector's evolution $|\psi (\tau )$$>$ to the energy momentum tensor field $T_{_{\mu \nu}}(x)$ is given by Equation (\ref{eq:25}); and the transformation from $T_{_{\mu \nu}}(x)$ to the metric field $h_{_{\mu \nu}}(x)$ follows by solving the linearised Einstein field equations. For weak gravitational fields, the relation between the tension scalar $R(x)$ and the metric field $h_{_{\mu \nu}}(x)$, and the relation between the term {\small$\sqrt{-g(x)}$$-$$1$} and the metric field, can be approximated by linear functions. 

\bigskip
\noindent
With the decomposition of the energy momentum tensor field $T_{_{\mu \nu}}(x)$ to the classical scenarios according to Equation (\ref{eq:27}) and the linear dependency between $R(x)$, {\small$\sqrt{-g(x)}$$-$$1$} and $T_{_{\mu \nu}}(x)$, we obtain the following decompositions:

\begin{equation}
\label{eq:35}
\begin{split}
R(x)=|c_{_{1}}|^{2} R_{_{1}}(x) + |c_{_{2}}|^{2} R_{_{2}}(x)  ~~~~~~~~~~~ \\
\mathsmaller{\sqrt{ - g(x)}}=|c_{_{1}}|^{2} \mathsmaller{\sqrt{ - g_{_{1}}(x)}} + |c_{_{2}}|^{2} \mathsmaller{\sqrt{ - g_{_{2}}(x)}}
\end{split}
\textrm{~~~~,~~~~~~~~~~~~~~~~~}
\end{equation}

~
\newline
\noindent where $R_{_{i}}(x)$ and {\small$\sqrt{-g_{_{i}}(x)}$} are respectively the tension scalars and the factors of the covariant volume elements of the classical scenarios following with the linear transformations of Equation (\ref{eq:34}) from the scenarios' energy momentum tensor fields $T_{_{\mu \nu i}}(x)$. For the Lagrangian density of the matter fields $L_{_{M}}(x)$, one obtains the same decomposition

\begin{equation}
\label{eq:36}
L_{_{M}}(x)=|c_{_{1}}|^{2} L_{_{M \, 1}}(x) + |c_{_{2}}|^{2} L_{_{M \, 2}}(x)
\textrm{~~,~~~~~~~~~~~~~~~~~}
\end{equation}

~
\newline
\noindent where the Lagrangian density is calculated analogously to the energy momentum tensor field in Equation (\ref{eq:25}) with a corresponding operator $\hat{L}_{_{M}}(x)$.

\bigskip
\noindent
By inserting the decompositions of Equations (\ref{eq:35}) and (\ref{eq:36}) into the Einstein-Hilbert action, we obtain, by a calculation similar to that of the total energy in Section \ref{sec:3.1} (Equation \ref{eq:17}), the following decomposition of the Einstein-Hilbert action according to the classical scenarios:

\begin{equation}
\label{eq:37}
S_{_{EH}}(\tau)=|c_{_{1}}|^{2}S_{_{EH1}}(\tau)+|c_{_{2}}|^{2}S_{_{EH2}}(\tau)+|c_{_{1}}|^{2}|c_{_{2}}|^{2}S_{_{G12}}(\tau)
\textrm{~~.~~~~~~~~~~~~~~~~~}
\end{equation}

~
\newline
\noindent Here $S_{_{EH1}}(\tau )$ and $S_{_{EH2}}(\tau )$ are respectively the Einstein-Hilbert actions of Classical Scenarios 1 and 2 alone, which are given by (cf. Equation \ref{eq:32}):

\begin{equation}
\label{eq:38}
S_{_{EH \, i}}(\tau)=\int^{\sigma (\tau)}_{...}\frac{d^{4}x}{c}\sqrt{-g_{_{i}}(x)}\left(\frac{R_{_{i}}(x)}{2\kappa} + L_{_{M \, i}}(x)  \right)
\textrm{\textsf{~~,~~~~~~~~~~~~~~~~~}}
\end{equation}

~
\newline
\noindent and the action $S_{_{G12}}(\tau )$ is given by

\begin{equation}
\label{eq:39}
S_{_{G12}}(\tau)=\int^{\sigma(\tau)}_{..}\frac{d^{4}x}{c}(\frac{R_{_{1}}(x)}{2\kappa} - \frac{R_{_{2}}(x)}{2\kappa} + L_{_{M \,1}}(x)-L_{_{M \,2}}(x))(\sqrt{-g_{_{2}}(x)}-\sqrt{-g_{_{1}}(x)})
\textrm{\textsf{~.}}
\end{equation}

~
\newline
\noindent This expression can be simplified as follows. The tension scalars of the classical scenarios $R_{_{i}}(x)$ can be expressed by their energy momentum tensor fields as follows: 

\begin{equation}
\label{eq:40}
R_{_{i}}(x)=\kappa T_{_{i}}(x)
\textrm{\textsf{~~,~~~~~~~~~~~~~~~~~~~~~~}}
\end{equation}

~
\newline
\noindent where $T_{_{i}}(x)$ are the contractions of these fields ($T_{_{i}}(x)$$\mathsmaller{\equiv}$$T^{\mu}_{~ \,  \mu i}(x)$). Equation (\ref{eq:40}) arises from a contraction of Einstein's field equation \cite{Gen-7}. Since the Lagrangian density of a free fermion is zero
\footnote{   
This follows with $L$$=$$c\bar{\psi}(x)[i\hbar\gamma^{\mu}\partial_{_{\mu}}$$-$$mc]\psi (x)$ and Dirac's equation $i\hbar \gamma^{\mu}\partial_{_{\mu}}\psi (x)$$-$$mc\psi (x)$$=$$0$ \cite{Gen-3}.
},
the Lagrangian densities $L_{_{M\, i}}(x)$ can be neglected ($|L_{_{M\, i}}(x)|$$<$$<$ $T_{_{i}}(x)$$\approx$$c^{2}\rho_{_{i}}(x)$), which leads to

\begin{equation}
\label{eq:41}
\begin{split}
S_{_{G12}}(\tau)=\frac{1}{2}\int^{\sigma(\tau)}_{..}\frac{d^{4}x}{c}(T_{_{1}}(x)-T_{_{2}}(x))(\sqrt{-g_{_{2}}(x)}-\sqrt{-g_{_{1}}(x)})
\textrm{\textsf{~~.~~~~~~~~~~~~~~~~~}}  \\
 \textrm{\textsf{~~~~~~~~~~~~~~~~~~~~~~~~~~~~~~~~~~~~~~~~~~~~~~~~~~~~~~~~~\footnotesize \textit{competition action (relativistic)}}}
\end{split}
\end{equation}

~
\newline
\noindent In the Newtonian limit, this expression passes over into our competition action according to Equation (\ref{eq:28}). This follows with the following approximations

\begin{equation}
\label{eq:42}
\sqrt{-g(x)}\approx 1+\frac{\Phi (x)}{c^{2}}
\textrm{~~,~~~~~~~~~~~}
T(x)\approx c^{2}\rho(x)
\textrm{~~~~~~~~~~~~~~~}
\end{equation}

~
\newline
\noindent for the Newtonian limit. Equation (\ref{eq:41}) is the relativistic generalisation of the competition action. The relativistic equivalent of the difference of the classical scenarios' mass distributions {\small$\rho_{_{1}}(\mathbf{x})$$-$$\rho_{_{2}}(\mathbf{x})$} in Equation (\ref{eq:28}) is the difference of the contracted energy momentum tensor fields {\small$T_{_{1}}(x)$$-$$T_{_{2}}(x)$}, and the relativistic equivalent of the difference of the classical scenarios' gravitational potentials {\small $\Phi_{_{2}}(\mathbf{x})$$-$$\Phi_{_{1}}(\mathbf{x})$} the difference of the factors of the covariant volume elements {\small$\sqrt{-g_{_{2}}(x)}$$-$$\sqrt{-g_{_{2}}(x)}$}, which are functions of the scenarios' metric fields $h_{_{\mu \nu i}}(x)$ ($g_{_{i}}(x)$$\mathsmaller{\equiv}$$det(\eta_{_{\mu \nu}}$$+$$h_{_{\mu \nu i}}(x))$). Equation (\ref{eq:41}), for the competition action between classical scenarios, is evidently covariant.

\bigskip
\noindent
From the decomposition of the Einstein-Hilbert action according to classical scenarios (Equation \ref{eq:37}), it follows that the competition action $S_{_{G12}}(\tau )$ can be interpreted similarly to the Di\'{o}si-Penrose energy in Section \ref{sec:3.1} as a measure of how much the preferred spacetime geometries of Classical Scenario 1 and 2 differ from each other, or of how strong the classical scenarios compete for spacetime geometry.

\bigskip
\bigskip
%
%
%
\subsection{Relativistic generalisation of the states' decay rates}                 
%
%
\label{sec:4.5}
In this section, we derive the relativistic generalisation of the states' decay rates, with which we derived the Di\'{o}si-Penrose criterion and Born's rule in Section \ref{sec:3.3}.

\bigskip   
\bigskip
\bigskip
\noindent
\textbf{Relativistic generalisation of classical scenarios' detuning actions} \\  
\noindent
We first derive the relativistic expressions for the classical scenarios' detuning actions (Equation \ref{eq:29}), with which the states' decay trigger rates can be expressed according to Equation (\ref{eq:31}). 

\bigskip
\noindent
We define the detuning action of a classical scenario by how much its Einstein-Hilbert action increases due to sharing of spacetime geometry with the other scenario. This definition will pass in the Newtonian limit into our former definition (Equation \ref{eq:29}) integrating the energy increase over time. How much a classical scenario's Einstein-Hilbert action increases due to sharing of spacetime geometry can be determined by a calculation similar to that of the states' energy increases from the superpositions total energy in Section \ref{sec:3.2}. The total Einstein-Hilbert action of e.g. Classical Scenario 1 is given by the terms in the superposition's Einstein-Hilbert action (Equation \ref{eq:37}), which are proportional to its intensity $|c_{_{1}}|^{2}$, which are $|c_{_{1}}|^{2}S_{_{EH1}}(\tau )$ and $|c_{_{1}}|^{2}|c_{_{2}}|^{2}S_{_{G12}}(\tau )$, and by normalising these terms to this intensity, which yields a total Einstein-Hilbert action of {\small$S_{_{EH1}}(\tau )$$+$$ |c_{_{2}}|^{2}S_{_{G12}}(\tau)$}. In the same way, the total Einstein-Hilbert action of Classical Scenario 2 yields {\small$S_{_{EH2}}(\tau )$$+$$ |c_{_{1}}|^{2}S_{_{G12}}(\tau)$}. The increases of the Einstein-Hilbert actions of Classical Scenarios 1 and 2 with respect to the case that they are alone and must not share spacetime geometry with the other are given by

\begin{equation}
\label{eq:43}
\begin{split}
S_{_{G1}}(\tau)=|c_{_{2}}|^{2}S_{_{G12}}(\tau) \\
S_{_{G2}}(\tau)=|c_{_{1}}|^{2}S_{_{G12}}(\tau) 
\end{split}
\textrm{\textsf{~~~~.~~~~~~~\footnotesize \textit{detuning actions (relativistic)}}}
\end{equation}

~
\newline
\noindent This relativistic version of the classical scenarios' detuning actions passes in the Newtonian limit over into our former definition according to Equation (\ref{eq:29}). This follows with the competition action for the Newtonian limit (Equation \ref{eq:28}) and Equation (\ref{eq:19}). 

\bigskip   
\bigskip
\bigskip
\noindent
\textbf{Relativistic generalisation of states' decay rates} \\  
\noindent
With the physical illustration found for the states' decay rates according to Equation (\ref{eq:31}) that the decay probability of a state $i$ $dp_{_{i\downarrow}}$ during a period of time is given by the increase of its detuning action $dS_{_{Gi}}$ divided by Planck's quantum of action $\hbar$, the relativistic generalisation of Equation (\ref{eq:31}) is: 

\begin{equation}
\label{eq:44}
\begin{split}
\frac{dp_{_{1\downarrow}}}{d\tau}=\frac{\frac{d}{d\tau}S_{_{G1}}(\tau)}{\hbar}  \\
\frac{dp_{_{2\downarrow}}}{d\tau}=\frac{\frac{d}{d\tau}S_{_{G2}}(\tau)}{\hbar} 
\end{split}
\textrm{\textsf{~~.~~~~~\footnotesize \textit{decay rates of states (relativistic)}}}
\end{equation}

~
\newline
\noindent where the system's state is followed up on an arbitrarily chosen hypersurface sequence $\sigma (\tau )$ (cf. Section \ref{sec:4.1}), and the detuning action $S_{_{Gi}}(\tau )$ is calculated up to this hypersurface. According to this result, the decay probability of a state $i$ $dp_{_{i\downarrow}}$ during the hypersurface moving from $\sigma (\tau )$ to $\sigma (\tau $$+$$d\tau )$ is given by the increase of its detuning action $dS_{_{Gi}}$ during this interval divided by $\hbar$.

\bigskip
\noindent
Equation (\ref{eq:44}) has no physical application in the context of our analysis of semiclassical gravity. However, it will obtain one in the Dynamical Spacetime approach, as we will see later.

\newpage
%
%
%
\section{Superpositions of more than two states}                 
%
%
\label{sec:5}
In this section, we extend our analysis of semiclassical gravity to superpositions of more than two states. The analysis of typical experiments shows that different states or classical scenarios often have identical mass distributions on some areas, which we combine to so-called \textit{local bundles} of states or classical scenarios on these areas. To measure how much the preferred spacetime geometries of local bundles of states or classical scenarios differ from each other on the regarded area, we introduce so-called \textit{local Di\'{o}si-Penrose energies} and \textit{local competition actions}. 

\bigskip
\noindent
In Section \ref{sec:5.1}, we define the local bundles, local Di\'{o}si-Penrose energies and local competition actions. In Section \ref{sec:5.2}, we calculate energy increases, detuning actions and decay rates of local bundles. In Section \ref{sec:5.3}, we apply the results to typical quantum mechanical experiments, and show why these experiments behave in accordance with Born's rule.

\bigskip
%
%
%
\ \subsection{Local bundles, Di\'{o}si-Penrose energies and competition actions }                 
%
%
\label{sec:5.1}
In this section, we define local bundles, local Di\'{o}si-Penrose energies and local competition actions.

%
\begin{figure}[b]
\centering
\includegraphics[width=12cm]{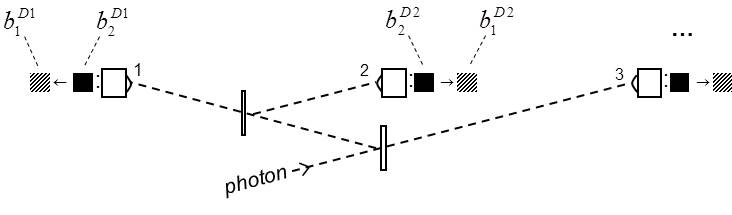}\vspace{0cm}
\caption{\footnotesize
Single-photon experiment generating a superposition of three states. 
}
\label{fig5}
\end{figure}

\bigskip   

\noindent
\textbf{Local bundles of states and classical scenarios} \\  
\noindent
The analysis of typical experiments shows that different states often have identical mass distributions on some areas. In the three-detector experiment in Figure \ref{fig5}, which generates a superposition of three states, where each state corresponds to a photon detection in one of the detectors, States 2 and 3 have on the area of Detector 1 identical mass distributions, since they both correspond to the case that the photon is not detected by Detector 1. The same applies for States 1 and 3 on the area of Detector 2, etc. This leads us to the following definition of \textit{local bundles}:

\begin{quote}
{\small
When several states or classical scenarios have identical wavefunctions and prefer identical spacetime geometries on an area $A$, they are a \textit{local bundle} $b^{A}_{\kappa}$ on $A$. 
}
\end{quote}

\noindent Here $\kappa$ is the \textit{bundle index}, for which we use Greek letters, and which is needed to distinguish several bundles on the same \textit{bundle area} $A$. Two states or classical scenarios $i$ and $j$ have identical wavefunctions on $A$, when the parts of the state vectors referring to $A$ are identical ($|\psi_{_{i}}$$>_{_{A}}$$=$$|\psi_{_{j}}$$>_{_{A}}$). This assumes that the state vector can be decomposed into a part referring to the bundle area $A$ and a part referring to the area outside of $A$ as {\small$|\psi_{_{i}}$$>$$=$$\psi_{_{i}}$$>_{_{A}}$$\otimes$$|\psi_{_{i}}$$>_{\neg A}$}. Two classical scenarios prefer identical spacetime geometries on $A$, when the metric fields $h_{_{\mu \nu i}}(x)$ resulting from their energy momentum tensor fields $T_{_{\mu \nu i}}(x)$ (by solving the linearised Einstein field equations) are identical on $A$. In the Newtonian limit, two states prefer identical spacetime geometries, when the gravitational potentials $\Phi_{_{i}}(\mathbf{x})$ resulting from their mass distributions $\rho_{_{i}}(\mathbf{x})$ are identical. The intensity of a local bundle $\kappa$ is defined by the sum over the intensities of its states or classical scenarios as

\begin{equation}
\label{eq:45}
|c_{_{\kappa}}|^{2}=\sum_{i\in b^{A}_{\kappa}}|c_{_{i}}|^{2}
\textrm{\textsf{~~.~~~~~~~~~~~~~~\footnotesize \textit{intensities of local bundles}}}
\end{equation}

~
\newline
\noindent For the three-detector experiment in Figure \ref{fig5}, we find three bundle areas corresponding to the areas of the three detectors. On the area of Detector 1, we  have the local bundle $b^{D1}_{1}$$=$$\{ 1\}$ consisting of State 1 only, and the local bundle $b^{D1}_{2}$$=$$\{ 2,3\}$ consisting of States 2 and 3, where the local bundle $b^{D1}_{1}$ corresponds to a photon detection, and the local bundle $b^{D1}_{2}$ to no photon detection. On the areas of Detectors 2 and 3 we obtain two local bundles accordingly, as shown in the figure.

\bigskip   
\bigskip
\bigskip
\noindent
\textbf{Local Di\'{o}si-Penrose energies} \\  
\noindent
Between two local bundles $b^{A}_{\kappa}$ and $b^{A}_{\nu}$ on the same bundle area $A$, we can define a \textit{local Di\'{o}si-Penrose energy} by restricting the integration in Equation (\ref{eq:4}) to the bundle area $A$:

\begin{equation}
\label{eq:46}
E^{^{A}}_{G\kappa \nu}=\frac{1}{2} \int_{\mathbf{x} \in A}  d^{3}\mathbf{x}(\rho_{_{\kappa}}(\mathbf{x})-\rho_{_{\nu}}(\mathbf{x}))(\Phi_{_{\nu}}(\mathbf{x})-\Phi_{_{\kappa}}(\mathbf{x}))
\textrm{\textsf{~~,~~~\footnotesize \textit{local Di\'{o}si-Penrose energy}}}
\end{equation}

~
\newline
\noindent where $\rho_{_{\kappa}}(\mathbf{x})$, $\rho_{_{\nu}}(\mathbf{x})$ are the mass distributions, and $\Phi_{_{\kappa}}(\mathbf{x})$, $\Phi_{_{\nu}}(\mathbf{x})$ the gravitational potentials of bundle $\kappa$ and $\nu$ respectively. The local Di\'{o}si-Penrose energy $E^{^{A}}_{G\kappa \nu}$ defines a measure of how much the preferred spacetime geometries of bundles $\kappa$ and $\nu$ differ from each other on the bundle area $A$. 

\bigskip
\noindent
When we have a superposition of more than two states for which we can describe the competition of its states for spacetime geometry with the help of our local Di\'{o}si-Penrose energies, as in the experiment of Figure \ref{fig5}, we obtain by a calculation similar to that in Section \ref{sec:3.1} (deriving Equation \ref{eq:17} for the total energy of a two-state superposition) the following total energy:

\begin{equation}
\label{eq:47}
E=\sum_{i}|c_{_{i}}|^{2}E_{_{i}} ~ + ~ \sum_{A}\sum_{\kappa < \nu} |c_{_{\kappa}}|^{2}|c_{_{\nu}}|^{2}E^{^{A}}_{G \kappa \nu}
\textrm{\textsf{~~,~~~~~~~~~~~~~~~~~~~~~~~~}}
\end{equation}

~
\newline
\noindent where $E_{_{i}}$ are the total energies of the single states (cf. Equation \ref{eq:18}). The outer sum of the second term runs over all bundle areas $A$, and the inner sum over all bundle pairs $\kappa$, $\nu$ on $A$.

\bigskip   
\bigskip
\bigskip
\noindent
\textbf{Local competition actions} \\  
\noindent
Similar to the local Di\'{o}si-Penrose energy between local bundles of states, we can define a local competition action between local bundles of classical scenarios by restricting the integration in Equation (\ref{eq:41}) to the bundle area $A$:

\begin{equation}
\label{eq:48}
\begin{split}
S^{^{A}}_{G \kappa \nu}(\tau)=\frac{1}{2}\int^{\sigma(\tau)}_{\stackrel{..}{~ x \in A}}\frac{d^{4}x}{c}(T_{_{\kappa}}(x)-T_{_{\nu}}(x))(\sqrt{-g_{_{\nu}}(x)}-\sqrt{-g_{_{\kappa}}(x)})
\textrm{\textsf{~~,~~~~~~~~~~~~~~~~~}}  \\
 \textrm{\textsf{~~~~~~~~~~~~~~~~~~~~~~~~~~~~~~~~~~~~~~~~~~~~~\footnotesize \textit{local competition action (relativistic)}}}
\end{split}
\end{equation}

~
\newline
\noindent where the contracted energy momentum tensor fields $T_{_{\kappa}}(x)$ and the factors of the covariant volumes elements {\small$\sqrt{-g_{_{\kappa}}(x)}$} of the local bundles result from their energy momentum tensor fields and metric fields by $T_{_{\kappa}}(x)$$=$$T^{\mu}_{~ \,  \mu \kappa}(x)$ and $g_{_{\kappa}}(x)$$=$$det(\eta_{_{\mu \nu}}$$+$$h_{_{\mu \nu \kappa}}(x))$. The  local competition action $S^{^{A}}_{G\kappa \nu}(\tau )$ defines a measure of how much the preferred spacetime geometries of bundles of classical scenarios $\kappa$ and $\nu$ differ from each other on the spacetime region, which is defined by the bundle area $A$, and limited towards the future by the hypersurface $\sigma (\tau )$.

\bigskip
\noindent
In the Newtonian limit, the local competition action between local bundles of classical scenarios can be calculated with the local Di\'{o}si-Penrose energies between the scenarios' states as follows (cf. Equation \ref{eq:28}):

\begin{equation}
\label{eq:49}
S^{^{A}}_{G \kappa \nu}(t)= \int^{t}_{..} dt E^{^{A}}_{G \kappa \nu}(t)
\textrm{\textsf{~~.~~~~~~\footnotesize \textit{local competition action (Newtonian limit)}}}
\end{equation}

~
\newline
\noindent Similar to Equation (\ref{eq:47}) for total energy, we can calculate the superposition's total Einstein-Hilbert action, which yields

\begin{equation}
\label{eq:50}
S_{_{EH}}(\tau)=\sum_{i}|c_{_{i}}|^{2} S_{_{EHi}}(\tau) ~ + ~ \sum_{A}\sum_{\kappa < \nu} |c_{_{\kappa}}|^{2}|c_{_{\nu}}|^{2}S^{^{A}}_{G \kappa \nu}(\tau)
\textrm{\textsf{~~,~~~~~~~~~~~~~~~~~~~}}
\end{equation}

~
\newline
\noindent where $S_{_{EHi}}(\tau )$ are the Einstein-Hilbert actions of the single classical scenarios alone (cf. Equation \ref{eq:38}).

\newpage
%
%
%
\subsection{Energy increases, detuning actions and decay rates of local bundles}                 
%
%
\label{sec:5.2}
In this section, we calculate for the local bundles energy increases, detuning actions and decay rates.

\bigskip   
\bigskip
\bigskip
\noindent
\textbf{Energy increases of local bundles} \\  
\noindent
The energy increase of a local bundle of states $b^{A}_{\kappa}$ occurring due to the sharing of spacetime geometry with the other bundles on the same bundle area $A$ can be determined similarly to the calculation of the states' energy increases in Section \ref{sec:3.2} from the superpositions total energy by identifying the terms in Equation (\ref{eq:47}), which are proportional to the bundle's intensity $|c_{_{\kappa}}|^{2}$, and by normalising these terms to this intensity. This yields

\begin{equation}
\label{eq:51}
E^{^{A}}_{G \kappa}=\sum_{\nu \neq \kappa}|c_{_{\nu}}|^{2}E^{^{A}}_{G \kappa \nu}
\textrm{\textsf{~~.~~~~~~~~~~~~~~\footnotesize \textit{energy increases of local bundles}}}
\end{equation}

~
\newline
\noindent This result is the intuitively expected generalisation of Equation (\ref{eq:19}). The energy increase of a local bundle $\kappa$ on $A$ depends on the intensities of the competing bundles $\nu$ on $A$ multiplied by the local Di\'{o}si-Penrose energy $E^{^{A}}_{G\kappa \nu}$ between the bundles. 

\bigskip   
\bigskip
\bigskip
\noindent
\textbf{Detuning actions of local bundles} \\  
\noindent
In the same way, we obtain the detuning action of a local bundle of classical scenarios $b^{A}_{\kappa}$ as follows:

\begin{equation}
\label{eq:52}
S^{^{A}}_{G \kappa}(\tau)=\sum_{\nu \neq \kappa}|c_{_{\nu}}|^{2}S^{^{A}}_{G \kappa \nu}(\tau)
\textrm{\textsf{~~,~~~~~~~~~~~~~~\footnotesize \textit{detuning actions of local bundles}}}
\end{equation}

~
\newline
\noindent and which is the intuitively expected generalisation of Equation (\ref{eq:43}). 

\bigskip   
\bigskip
\bigskip
\noindent
\textbf{Decay rates of local bundles} \\  
\noindent
The decay rate of a local bundle follows similarly to the calculation of the states' decay rates in Section \ref{sec:3.3} by dividing its energy increase $E^{^{A}}_{G\kappa}$ by Planck's constant $\hbar$ as follows:

\begin{equation}
\label{eq:53}
\frac{dp^{^{A}}_{\kappa \downarrow}}{dt} = \frac{E^{^{A}}_{G \kappa}}{\hbar}
\textrm{\textsf{~~.~~~~~~~~~~~~~~\footnotesize \textit{decay rates of local bundles (Newtonian limit)}}}
\end{equation}

~
\newline
\noindent The relativistic generalisation of a local bundle's decay rate follows with Equation (\ref{eq:44}) and the detuning action $S_{_{G\kappa}}(\tau )$ of the local bundle as:

\begin{equation}
\label{eq:54}
\frac{dp^{^{A}}_{\kappa \downarrow}}{d\tau} = \frac{\frac{d}{d\tau}S^{^{A}}_{G \kappa}(\tau)}{\hbar}
\textrm{\textsf{~~.~~~~~~\footnotesize \textit{decay rates of local bundles (relativistic)}}}
\end{equation}

~
\newline
\noindent This result has the same physical illustration as Equation (\ref{eq:44}). The decay probability $dp^{^{A}}_{\kappa\downarrow}$ of a local bundle $b^{A}_{\kappa}$ during the hypersurface moving from $\sigma (\tau )$ to $\sigma (\tau $$+$$d\tau )$ is given by the increase of the bundle's detuning action $dS^{^{A}}_{G\kappa}$ during this interval divided by $\hbar$.

%
\begin{figure}[b]
\centering
\includegraphics[width=7cm]{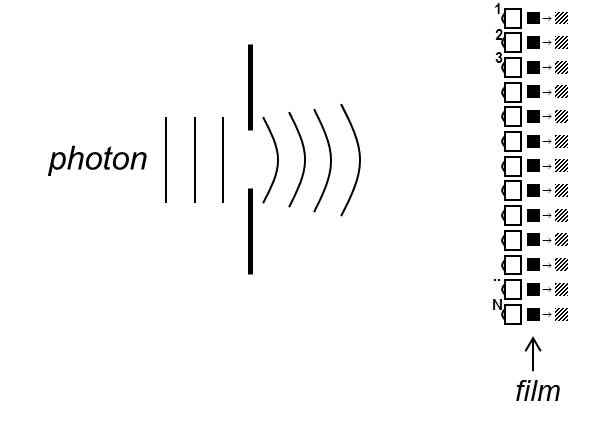}\vspace{0cm}
\caption{\footnotesize
Measurement of a photon with a film that can be modelled with a large number of small mass displacing detectors.
}
\label{fig6}
\end{figure}

\bigskip
\bigskip
%
%
%
\ \subsection{Derivation of Born's rule for typical quantum mechanical  experiments}                 
%
%
\label{sec:5.3}
In this section, we apply the results of Sections \ref{sec:5.1} and \ref{sec:5.2}, and show how our derivation of Born's rule for two-state superpositions in Section \ref{sec:3.3} can be generalised for typical quantum mechanical experiments. Our analysis will show that these experiments have a property in common: they have never more than two local bundles on one bundle area. With the help of this property, we can show why all experiments performed so far behave in accordance with Born's rule. The question of whether cases with more than two local bundles lead to different behaviours will be discussed in the Dynamical Spacetime approach in Part 2 \cite{P2}.

\bigskip
\bigskip
\noindent
Typical quantum mechanical experiments can be categorised into two groups. Experiments with active measuring devices, such as the three-detector experiment in Figure \ref{fig5}, and experiments with passive measuring devices, such as e.g. films or cloud chambers. Both groups can be discussed in a common picture by modelling passive measuring devices with a large number of small mass displacing detectors, as shown in Figure \ref{fig6} for a photon measurement with a film. 

\bigskip
\noindent
If one analyses typical quantum mechanical experiments from the local bundles' point of view, one finds that they generate never more than two local bundles on one bundle area, which refers to the cases that the particle "is", or "is not", detected by the detector on the bundle area, as one can see with Figures \ref{fig5} and \ref{fig6}. With this common property of these experiments, we can extend our derivation of Born's rule for two-state superpositions in Section \ref{sec:3.3} to the typical quantum mechanical experiments.

\bigskip
\noindent
To discuss reduction probabilities with the decay rates of local bundles (derived in Section \ref{sec:5.2}), we have to make one assumption that can be derived from the Dynamical Spacetime approach in Part 2 \cite{P2}. It is that the two competing bundles on a bundle area $A$, $b^{A}_{1}$ and $b^{A}_{2}$, decay in favour of each other. This means that the bundle $b^{A}_{1}$, whose decay is described by the decay rate $dp^{^{A}}_{1\downarrow}/dt$ according to Equation (\ref{eq:53}), decays to bundle $b^{A}_{2}$ and vice versa.

\bigskip
\noindent
The local bundle on the bundle area of detector $i$ corresponding to the case that the particle is detected by this detector (the "detection state") consists of the single state $i$; and the local bundle corresponding to the case that the particle is not detected by this detector (the "no-detection state") consists of the superposition of all states except $i$, which we label $\neg  i$. The local bundles of the detection and no-detection state are labelled $b^{Di}_{i}$ and $b^{Di}_{\neg i}$. The decay rates of these bundles follow with Equations (\ref{eq:53}) and (\ref{eq:51}) to be:

\begin{equation}
\label{eq:55}
\begin{split}
\frac{dp^{^{Di}}_{i \downarrow}}{dt } =\sum_{\mathsmaller{j\neq i}}|c_{_{j}}|^{2} \frac{E^{^{Di}}_{G}}{\hbar} \\
\frac{dp^{^{Di}}_{\neg i \downarrow}}{dt } =~~~~~~ |c_{_{i}}|^{2} \frac{E^{^{Di}}_{G}}{\hbar}
\end{split}
\textrm{\textsf{~~~~,~~~~~~~~~~~~~~~~~}}
\end{equation}

~
\newline
\noindent where we abbreviated the local Di\'{o}si-Penrose energy between the detection and no-detection states $E^{^{Di}}_{Gi \neg i}$ by $E^{^{Di}}_{G}$. The decay rates are, as expected from Born's rule, proportional to the intensity of the competing bundle in favour of which the bundle decays, which is $\sum_{\mathsmaller{j\neq i}}|c_{_{j}}|^{2}$ for the superposition $\neg  i$ and $|c_{_{i}}|^{2}$ for state $i$.

\bigskip
\bigskip
\noindent
With Equation (\ref{eq:55}), we can discuss how the photon in Figure \ref{fig6} localises when it is measured with the film. The initial superposition of $N$ states ($N$=number of detectors) decays with a low probability directly to a state $i$, and a high one to a superposition $\neg  i$ consisting of $N$$-$$1$ states. In the latter case, the superposition $\neg  i$ decays accordingly with a low probability to a single state, and a high one to a superposition of $N$$-$$2$ states. This procedure repeats in subsequent decays, until the superposition has reduced to a single state $i$. The initial superposition of $N$ states can decay in many different ways to the same final state $i$. From Equation (\ref{eq:55}), i.e. the decay probability to a bundle is proportional to the bundle's intensity, one can easily see that the probability of all ways leading to a final state $i$ follows Born's rule, i.e.

\begin{equation}
\label{eq:56}
p_{_{i}} = |c_{_{i}}|^{2}
\textrm{\textsf{~~.~~~~~~~~~~~~~~~~~~~\footnotesize \textit{Born's rule}}}
\end{equation}

~
\newline
\noindent

\newpage
%
%
%
\section{Discussion}                 
%
%
\label{sec:6}
This work has shown how the characteristic energy of the Di\'{o}si-Penrose criterion is related to semiclassical gravity, and that the exact factor $\xi$ of this energy follows from the approaches of Di\'{o}si and Penrose and from semiclassical gravity to be $\xi$$=$$\frac{1}{2}$. The fact that the Di\'{o}si-Penrose criterion also has a relation to semiclassical gravity underlines its generality for gravity-based collapse models. 

\bigskip
\noindent
Our analysis of semiclassical gravity demonstrated that the sharing of spacetime geometry leads to energy increases of the states, which scale with the Di\'{o}si-Penrose energy, and which are proportional to the intensities of the states competing with the regarded one for spacetime geometry. These energy increases followed from the fact that the states reside in semiclassical gravity in the superposition's mean gravitational potential.

\bigskip
\noindent
It was shown how the spacetime analogue of the Di\'{o}si-Penrose energy, the competition action, can be relativistically generalised with the help of the Einstein-Hilbert action. With this result, we will be able to formulate the Dynamical Spacetime approach in a covariant form in Part 2 \cite{P2}. Whether this result is also helpful for the still-missing relativistic generalisations of the approaches of Di\'{o}si and Penrose is still open. 

\bigskip
\noindent
Our analysis of semiclassical gravity has also shown that there could be a relation between semiclassical gravity and Born's rule. This relation resulted from the property found that all typical quantum mechanical experiments performed so far lead to never more than two local bundles (i.e. two different mass distributions) at one location, and that the energy increases, respectively decay rates, of the locally competing bundles are proportional to the intensity of the competing bundle in favour of which they decay. Whether a derivation of Born's rule from a physical argument, such as semiclassical gravity, makes sense in the light of the fact that there are several derivations of Born's rule deriving it from fundamental principles, such as those of Gleason \cite{Born-6}, Deutsch \cite{Born-7}, Zurek \cite{Born-8,BornNS-1} and others \cite{Born-2,Born-5}, can only be discussed in the context of a concrete collapse model, such as the Dynamical Spacetime approach, which will be given in Part 2 \cite{P2}.

\bigskip
\bigskip
\noindent
The main purpose of this paper was to prepare, via our analysis of semiclassical gravity, the derivation of the Dynamical Spacetime approach to wavefunction collapse in Part 2 \cite{P2}. The concepts derived here, such as the competition and detuning actions, the classical scenarios and the decay rates of states and local bundles, will be used for the derivation of the Dynamical Spacetime approach in Part 2 \cite{P2}. The Dynamical Spacetime approach enhances semiclassical gravity and enables an explanation of wavefunction collapse by postulating that the spacetime region on which quantum fields exist and on which the wavefunction's evolution can be regarded is bounded towards the future by a spacelike hypersurface, which is dynamically expanding towards the future. Collapse is displayed in the way that the wavefunction's evolution becomes unstable at certain critical expansions of spacetime, at which it reconfigures via a self-reinforcing mechanism quasi-abruptly to an evolution resembling a classical trajectory, which can be described with our classical scenarios. The critical expansion of spacetime, at which the wavefunction's evolution becomes unstable for reconfiguration, can be determined with the competition action, which is determined on the available spacetime region. A superposition of two classical scenarios collapses when the competition action between them coincides with Planck's quantum of action. The probabilities with which the wavefunction's evolution reconfigures at the critical expansions of spacetime to one of the classical scenarios depend on smallest intensity fluctuations of the states, which can be described with the help of the decay rates of the states and local bundles described here. 

\bigskip   
\bigskip
\bigskip
\bigskip
\noindent
\textbf{\small Acknowledgements} \\  
\noindent
{\small
I would like to thank my friend Christoph Lamm for supporting me and for proofreading the manuscript.
}

%
%
%

\end{document}